\documentclass[a4paper,12pt]{article}
\usepackage[left=2cm,right=2cm]{geometry}
\usepackage{epsf}
\usepackage{epsfig,float}
\usepackage{a4wide,graphicx}
\begin{document}
\title{Charm and hidden charm scalar mesons in the nuclear medium}

\author{R. Molina$^1$, D. Gamermann$^1$, E. Oset$^1$ and L. Tolos$^{2,3}$}
\maketitle

\begin{center}
$^1$ Departamento de F\'{\i}sica Te\'orica and IFIC,
Centro Mixto Universidad de Valencia-CSIC,
Institutos de Investigaci\'on de Paterna, Aptdo. 22085, 46071 Valencia, Spain\\
$^2$ Frankfurt Institute for Advanced Studies, Goethe-Universit\"at Frankfurt am Main, Ruth-Moufang-Str. 1, 60438 Frankfurt am Main, Germany. \\
$^3$ Theory Group, KVI, University of Groningen, Zernikelaan 25, 9747 AA Groningen, The Netherlands

\end{center}

\date{}

\abstract{ We study the renormalization of the properties of low lying
charm and hidden charm scalar mesons in a nuclear medium, concretely of the 
$D_{s0}(2317)$ and the theoretical hidden charm state  $X(3700)$.
We find that for the $D_{s0}(2317)$, with negligible width at zero density, the
width becomes about $100 ~{\rm MeV}$ at normal nuclear matter density, while in the
case of the $X(3700)$ the width becomes as large as $200~{\rm MeV}$.
We discuss the origin of this new width and trace it to reactions occurring in 
the nucleus, while offering a guideline for future experiments testing these changes. We also show how those medium modifications will bring valuable information on the nature of the scalar
resonances and the mechanisms of the interaction of $D$ mesons with nucleons and
nuclei. }

\section{Introduction}

The modification of the properties of elementary particles in nuclei is a rich field
that helps simultaneously to learn about excitation mechanisms in the nucleus as
well as properties of the elementary particles \cite{Post:2003hu}. Important
features about the nature of certain particles can be better observed in nuclei.
Take for instance the $\Lambda(1520)$. This resonance couples strongly to $\pi
\Sigma(1385)$, but this is hardly visible in the decay of the $\Lambda(1520)$
particle since there is no phase space for it. When this resonance is placed 
inside a nucleus the pion can become a particle hole excitation ($ph$) and one 
gains $140~{\rm MeV}$ of phase space for the decay, which results in a considerable 
increase in the width of the $\Lambda(1520)$ in the medium \cite{Kaskulov:2005uw}. Take another
example: the
$\omega$ meson. This meson decays into $3 \pi$, which is supposed to go through $\rho \pi$,
but there is no phase space for decay in this channel except for the $\rho$ 
width. Once again the $\pi \pi$ decay channel of the $\rho$ will be modified in
the nucleus, as well as the individual $\pi$, which can become again a $ph$
excitation producing a much larger phase space for the decay of the $\omega$. 
It is then spectacular to find that the width of the  $\omega$ in the medium at
nuclear matter density becomes as large as $100-150~{\rm MeV}$ 
\cite{kaskulov,omegawidth}. This discussion is only qualitative and more subtle details must be considered as done in other works \cite{Wolf,Broniowski,Broniowski2}.

Among so many examples in the literature, the case of the renormalization of the
properties of the scalar mesons has played a special role. One reason is that
there is a long debate on the nature of the scalar mesons,  with
different assumptions about their nature as $q \bar{q}$ states, $K \bar{K}$ 
molecules, mixtures of $q \bar{q}$ with meson-meson components, 
or dynamically generated resonances from the interaction of coupled
channels of two pseudoscalars \cite{Beveren}-\cite{kaiser}.
We will study here the properties in the medium of two dynamically generated states, meaning states made out of two mesons. Pioneer work in the study of dynamically generated resonances was done in \cite{Beveren, Tornqvist}, where starting from one seed of $q\bar{q}$, a large meson cloud was demanded to explain data of the low lying scalars. Further work in this direction followed using the unitary coupled channel chiral approach in \cite{npa,iam,kaiser}. In this sense, the predictions of
the different models on the medium modification of these resonances are
important in order to put stringent constraints on the different assumptions
about the nature of the resonances.

Among the low lying scalars, the $\sigma(600)$ has been the most studied.
Several theoretical approaches have predicted strong medium
effects on the $\pi \pi$ interaction in the scalar isoscalar ($\sigma$) channel.
In \cite{sigmaulf}, the $\sigma$ and $\pi$ mesons at high density are studied within the Nambu-Jona-Lasinio model and the authors find that the $m_\sigma$ drops subtantially with the density, whereas the $m_\pi$ increase at higher densities. The $\pi$ was previously studied at finite temperature and density within the same model in \cite{piulf}.
 In Ref.  \cite{Hatsuda:1999kd}, Hatsuda et al. studied the $\sigma$ propagator
in the linear $\sigma$ model and found an enhanced and narrow spectral 
function  near the $2\pi$ threshold  caused by the partial restoration of the
chiral symmetry, where $m_\sigma$ would approach $m_\pi$. The same
conclusions were reached using the nonlinear chiral Lagrangians  in Ref.
\cite{Jido:2000bw}.
Similar results, with large enhancements in the $\pi\pi$ amplitude around the 
$2\pi$ threshold, have been found in  quite different approaches by studying the
$s-$wave, $I=0$ $\pi\pi$ correlations in nuclear matter
\cite{pipi}. In these cases
the modifications of the $\sigma$ channel are induced by the strong $p-$wave
coupling of the pions to the particle-hole ($ph$) and $\Delta$-hole ($\Delta
h$) nuclear excitations.  A more recent and updated theoretical work on the
evolution of the $\sigma$ poles in the medium can be seen in
\cite{Cabrera:2005wz}.

On the experimental side, there are also several results showing strong medium
effects in the $\sigma$ channel at low invariant masses in the $A(\pi,2\pi)$ 
\cite{pidospi}
and $A(\gamma,2\pi)$  \cite{Messchendorp:2002au} reactions, which have been the
object of study in \cite{VicenteVacas:1999xx}, \cite{Roca:2002vd} and  
\cite{Muhlich:2004zj}.

 The $f_0(980)$ and $a_0(980)$ scalar mesons have also been analyzed in the
nuclear medium \cite{Oset:2000ev}, but unlike the case of the $\sigma$, no
experimental action has been taken in this case. 

    In the present paper we retake research along these lines and study the
medium modification of the scalar mesons which are dynamically generated in the 
charm sector. Concretely, we shall study the medium modification of the
$D_{s0}(2317)$, which is obtained in the theoretical studies of 
\cite{Kolomeitsev:2003ac}-\cite{Gamermann:2006nm}. 
 Within the context of lattice gauge theories, in \cite{NievesDs} the authors find hints that there is a $DK$ bound state that can be identified with the $D_{s0}(2317)$. In
addition, we shall also study the medium modifications of a hidden charm scalar
meson predicted in \cite{Gamermann:2006nm}, for which there are some indications 
that could have been observed in the experiment of the Belle collaboration
\cite{Abe:2007sy} through the reanalysis done in \cite{Gamermann:2007mu}. In this 
experiment a broad bump is seen in the $D\bar{D}$ mass distribution around the 
$D\bar{D}$ threshold, which in \cite{Gamermann:2007mu} was shown to be better 
explained by the $X(3700)$ pole of the $D\bar{D}$ amplitude below threshold than 
by the Breit Wigner distribution proposed in the experimental paper. However, it 
should be noted that a better fit to the data with a Breit Wigner distribution 
than the one of \cite{Abe:2007sy} was obtained in \cite{Gamermann:2007mu}, 
yet with a $\chi^2$ value slightly worse than the one obtained with the pole below
threshold.

\section{Brief discussion on the dynamical generation of the $D_{s0}(2317)$ and $X(3700)$}

We follow the details of \cite{Gamermann:2006nm}, where a Lagrangian is taken
 for the interaction of two pseudoscalar mesons. The Lagrangian is an 
 extrapolation to $SU(4)$ of the $SU(3)$ chiral Lagrangian used in
 \cite{npa,kaiser} 
 to generate the scalar mesons $\sigma(600)$, $f_0(980)$, $a_0(980)$ and 
 $\kappa(900)$ in the light sector, but with the $SU(4)$ symmetry strongly 
 broken, mostly due to the explicit consideration of the masses of the vector 
 mesons exchanged between pseudoscalars in the equivalent theory using the 
 hidden gauge formalism for the vector mesons \cite{Bando:1984ej}-\cite{ulfvec}. 
 A different breaking of $SU(4)$ is also considered in 
 \cite{Gamermann:2006nm}, following general rules of $SU(n)$ breaking 
 \cite{Walliser:1992vx}, which serves as an indication of theoretical 
 uncertainties. We follow here the version based on the hidden gauge formalism, 
 including the exchange of heavy (charmed) vector 
 mesons in the Lagrangian.
  
  We would like to put the $SU(4)$ breaking in a certain context. 
 The basic assumption underlying \cite{Gamermann:2006nm,Gamermann:2007mu} is
 that the vertices in the hidden gauge formalism, of four vectors or three 
 vectors, are approximately $SU(4)$ symmetric. As a first step, the kernel of the Bethe Salpeter equation (the potential) already breaks SU(4) symmetry in the terms that exchange heavy vectors, as we have mentioned. Later on, the amplitudes calculated
 with the unitary approach break $SU(4)$ symmetry because the physical masses of the 
 particles are used to respect exactly thresholds and unitarity in coupled channels. This situation is already present in $SU(3)$. The starting lowest-order chiral Lagrangians are $SU(3)$ symmetric and the amplitudes obtained break $SU(3)$ symmetry due to the different masses of the particles belonging to the same $SU(3)$ multiplet. As an example, one of the $\Lambda(1405)$ states \cite{Jido} and the $\Lambda(1670)$, which appear in the approach of \cite{Jido,bennhold}, are degenerate if the masses within the same $SU(3)$ multipletes are taken equal. One can see there an example of a large $SU(3)$ breaking present in nature, which, however, is consistent with assuming an exact $SU(3)$ meson-baryon interaction Lagrangian.
  
  With these assumptions for the SU(4) symmetry and its breaking one gets realistic results for the spectra of mesons in \cite{Gamermann:2006nm,Gamermann:2007mu}, which also agree with those obtained in the heavy quark formalism for the case of light-heavy meson interaction \cite{Guonew}, up to a mass term with no practical consequences. It is also interesting to note that the same basic $SU(4)$ assumptions are done for the interaction of mesons with baryons in \cite{lutzmed,mizutani,Tolos:2007vh} and one obtains realistic results concerning the $\Lambda_c(2595)$ and $\Sigma_c(2800)$ resonances.
  
  Following \cite{npa}, one derives the kernel (potential) from the lowest
  order Lagrangian and iterates it to generate all the terms of the Bethe 
  Salpeter series, which can be summed up in the on shell formalism 
  \cite{nsd,ollerulf} by means of
\begin{equation}
T=[1-V G]^{-1} V \ ,
\label{bethe}
\end{equation}
where $T$ is a matrix in the space of coupled channels representing the 
transition scattering amplitude from one channel to another and  $V$ the 
equivalent matrix for the transition potential. The diagonal matrix in 
the coupled channel space $G$ accounts for the loop integral of the two particle propagator of any intermediate state 
 \begin{eqnarray} 
 G_{ii}&=&i\int {dq^4\over (2\pi)^4} {1\over q^2-m_1^2+i\epsilon}{1\over (P-q)^2-m_2^2+i\epsilon}=
 \label{loopint}\\
 &&{1 \over 16\pi ^2}\biggr[ \alpha _i+Log{m_1^2 \over \mu ^2}+{m_2^2-m_1^2+s\over 2s}
  Log{m_2^2 \over m_1^2} \nonumber\\
  &+&{\bar p\over \sqrt{s}}\Big( Log{s-m_2^2+m_1^2+2\bar p\sqrt{s} \over -s+m_2^2-m_1^2+
  2\bar p\sqrt{s}}+Log{s+m_2^2-m_1^2+2\bar p\sqrt{s} \over -s-m_2^2+m_1^2+  2 \bar p\sqrt{s}}\Big)\biggr]
  \label{loopf} \ ,
\end{eqnarray}
where $P$ is the total four momentum, $q$ one of the pseudoscalar
four momentum,  $\bar  p$  the on shell three momentum and $m_1$, $m_2$ the masses of the two 
pseudoscalars. 

This integral requires dimensional regularization by means of a substraction
constant, $ \alpha _i$, or cutting the three dimensional integral in $G$ with a
cut off. Both methods establish equivalent schemes in a certain chosen region
of energies \cite{ollerulf}. However, the use of the dimensional regularization
method relies upon Lorentz covariance of magnitudes, which is lost in nuclei
where one has a privileged reference frame, the one where the nucleus is at
rest.  As a result, the use of the dimensional regularization method
introduces pathologies when including the selfenergy of the particles in the medium, which are avoided with the
use of a cut off \cite{laurap}. Hence, in the channels where we renormalize the particles in the medium, we stick to the cut off formalism. For all the other channels we use the same dimensional regularization approach of \cite{Gamermann:2006nm} which guarantees unitarity. The cut off method is used only in the channels $DK$ and $D\bar{D}$, and the values of the cut off will be shown later on, but they are much bigger than the on-shell three momenta of the particles in the loops, such that the imaginary part of $G_{ii}$ is obtained exactly in the free case. 
In free space we have
\begin{equation}
G_{ii}(s)=\int_0^{q_{max}} \frac{q^2 dq}{(2\pi)^2} \frac{\omega_1+\omega_2}{\omega_1\omega_2 [{(P^0)}^2-(\omega_1+\omega_2)^2+i\epsilon]   } \ ,\label{loopcut}
\end{equation}
where $q_{max}$ stands for the cut off, $\omega_i=(\vec{q}\,^2_i+m_i^2)^{1/2}$ and the center-of-mass energy ${(P^0)}^2=s$.  The expression of $G_{ii}$ will be changed to account for the medium effects on
the pseudoscalar mesons in  Sec.~\ref{two}.

 The $T$ matrix of Eq.~(\ref{bethe}) generates poles for some quantum numbers. We look for them in the second Riemann sheet for the channels which are open and those poles are associated to resonances. For the channels where we found bound states, i.e., states below  threshold, the poles appear in the first Riemann sheet. This is actually the case for  $X(3700)$ in the $D\bar{D}$ and also for the $D_{0s}(2317)$ in the $DK$ channel. Close to a pole,
the amplitude looks like 

\begin{equation}
T_{ij}\approx\frac{g_i g_j}{z-z_R} \ , 
\label{polos}
\end{equation}
where ${\rm Re}\, z_R$ gives the mass of the resonance 
and ${\rm Im}\, z_R$ the half width. The constants $g_i$, obtained from the residues of 
the amplitudes, provide the coupling of the resonance to a particular channel 
and indicate the relevance of this channel in building up the 
resonance. This said, it is useful to recall that the $X(3700)$  and $D_{s0}(2317)$
are obtained from poles of the scattering matrix with
 quantum numbers ($C=0$, $S=0$, $I=0$) and ($C=1$, $S=1$, $I=0$), respectively. In Tables 1 and 2, we show  
 the coupling of each state to the different channels 
 contributing to these quantum numbers, which are obtained from \cite{Gamermann:2006nm} but including
the $\eta'$ and the mixing with $\eta$ (see \cite{withzou}).

\begin{table}[h]
\begin{center}
\begin{tabular}{c|c|c|c}
\hline
Channel&Re($g_X$) [MeV]&Im($g_X$) [MeV]&$|g_X|$ [MeV]\\
\hline
\hline
$\pi^+\pi^-$&9&83&84\\
\hline
$K^+K^-$&5&22&22\\
\hline
$D^+D^-$&5962&1695&6198\\
\hline
$\pi^0\pi^0$&6&83&84\\
\hline
$K^0\overline{K^0}$&5&22&22\\
\hline
$\eta\eta$&1023&242&1051\\
\hline
$\eta\eta'$&1680&368&1720\\
\hline
$\eta'\eta'$&922&-417&1012\\
\hline
$D^0\overline{D^0}$&5962&1695&6198\\
\hline
$D_s^+D_s^-$&5901&-869&5965\\
\hline
$\eta_c\eta$&518&659&838\\
\hline
$\eta_c\eta'$&405&9&405\\
\hline
\end{tabular}
\caption{$X(3700)$: Couplings of the pole at (3722-$i$18) MeV to the channels (C=0, S=0,I=0).} \label{x3700}
\end{center}
\end{table}

\begin{table}[h]
\begin{center}
\begin{tabular}{c|c|c|c}
\hline
Channel&Re($g_{D_{s0}}$) [MeV]&Im($g_{D_{s0}}$) [MeV]&$|g_{D_{s0}}|$ [MeV]\\
\hline
\hline
$K^+D^0$&5102&0&5102\\
\hline
$K^0D^+$&5102&0&5102\\
\hline
$\eta D_s^+$&-2952&0&2952\\
\hline
$\eta' D_s^+$&4110&0&4110\\
\hline
$\eta_C D_s^+$&2057&0&2057\\
\hline
\end{tabular}
\caption{$D_{s0}(2317)$: Couplings of the pole at 2317 MeV to the channels (C=1, S=1,I=0).} \label{ds02317}
\end{center}
\end{table}

From these tables we observe the following:

1) The heavy singlet, hidden charm state $X(3700)$ couples most strongly to 
$D\bar{D}$. Next it couples to $D_s \bar{D}_s$.  However, from the square of couplings 
which would enter into the selfenergy of the resonance due to the intermediate meson-meson states, one gets a factor of two more weight for the $D\bar{D}$ than the 
$D_s \bar{D}_s$ states. On the other hand, while the pole around 3700 MeV is 
close to the $D\bar{D}$ threshold of  3738 MeV, it is about 238 MeV away 
from the $D_s \bar{D}_s$ threshold of 3938 MeV. The off shellness of the 
$D_s \bar{D}_s$ in the loop function of Eq.~(\ref{loopcut}) further reduces the 
contribution of this channel to less than 10\% of the $D\bar{D}$. This is 
important to note since, when dealing with the medium correction in the next 
sections, we shall consider the normalization of $D$ but not the one of $D_s$.

2) The $D_{s0}(2317)$ couples most strongly to $DK$. The next channels are the 
$D_s \eta'$ and $D_s \eta$. The same considerations as before lead to a similar relative 
contribution of $D_s \eta'$ and $D_s\eta$ with respect to the dominant $DK$ channel. Once again, 
we are lead to deal with the medium modification of the $D$ meson, those of the
$K$ state being relatively unimportant since the $KN$ interaction is  
not too strong and has no singularities \cite{tolos-dani}.

It should be noted that in the step from $SU(3)$ to $SU(4)$ we introduce more 
uncertainties than one has in $SU(3)$. That is the reason why in \cite{Gamermann:2006nm}
two different models, which break $SU(4)$ in different ways, were used to see the 
uncertainties in the results. It was found there that the results for the $X(3700)$ and $D_{s0}(2317)$ states were rather independent of the model, while other states predicted here were more 
model dependent. Even though, it is important to make an evaluation of the 
theoretical uncertainties for those states and, for this purpose, we have followed the same approach as in
\cite{Gamermann:2006nm}. Thus, we evaluate the results using a sample of parameters
which are varied within the reasonable limits discussed in \cite{Gamermann:2006nm}. In
particular we vary the $f_{\pi}$ and $f_{D}$ parameters in the range $f_{\pi}\in 
[85,\,115]~{\rm MeV}$ and $f_{D}\in [146,\,218]~{\rm MeV}$. This alone gives a good 
approximation to the uncertainties and we will perfom the medium calculations 
evaluating the uncertainties with this method.

\begin{table}
\begin{center}
\begin{tabular}{c|cc}
\hline
Channel & $|g_i|$ Model A [MeV]& $|g_i|$ Model B [MeV] \\
\hline
\hline
$DK$ & 7215 & 7503 \\
$D_s\eta$ & 2952 & 3005 \\
$D_s\eta\prime$ & 4110 & 4146 \\
$D_s\eta_c$ & 2058 & 1246 \\
\hline
\end{tabular}
\caption{Couplings of the $D_{s0}(2317)$ to its building blocks. Model A refers to the model using both, $f_\pi$ and $f_D$ in the couplings, while in Model B only $f_\pi$ is used, respecting constrains from chiral symmetry. The channels are in isospin basis. The position of the pole is fixed in both models to $2317$ MeV, taking $\alpha_H=-1.48$ in the model A, and $\alpha_H=-1.16$ in the model B ($\alpha_H$ means the substraction constant used in \cite{Gamermann:2006nm} for the channels involving at least one heavy pseudoscalar meson).}
\label{tab:newG}
\end{center}
\end{table}

The main results of the paper are the effects of the medium in the $D_{s0}(2317)$. The results for this resonance were obtained in \cite{Gamermann:2006nm}, where the prescription of using $f_\pi$ for the light mesons and $f_D$ for the heavy ones was used. For the  case of scattering of light mesons with heavy ones, chiral symmetry requires the use of $f_\pi$ in all cases (eventually $f_K$ if kaons are involved). For this reason we redo the calculations with the new chiral symmetric prescription. We also introduce the novelty with respect to \cite{Gamermann:2006nm} of considering also the $\eta'$ in the set of pseudoscalar mesons according to the method of \cite{withzou}. We should bear in mind that the substraction constant of Eq. (\ref{loopf}) was slightly tuned from its natural value to get the mass of the $D_{s0}(2317)$ at the right place. Hence, we do the same here and then we look at the results obtained for the couplings to the channels in the two cases. Those are shown in Table \ref{tab:newG}. As we can see, the results are very similar. For the most important building block, the $DK$ channel, the differences of the coupling are of $4\%$, indicating that the errors induced by the explicit chiral symmetry breaking of \cite{Gamermann:2006nm} are very small. Yet, in the present paper we shall use the chiral symmetric version described here. Using $f_K$ instead of $f_\pi$ leads to much smaller differences than those in Table \ref{tab:newG}. However, we shall use just $f_\pi$. 

For the $X(3700)$, which comes mostly from $D\bar{D}$, we have no such constraints from chiral symmetry and we follow the approach of \cite{Gamermann:2006nm}, except for the inclusion of the $\eta'$ channel. We evaluate uncertainties in the results, though, by using again $f_{\pi}$ instead of both $f_D$ and  $f_{\pi}$. The stability of the couplings with respect to the changes done can be partly justified by recalling that, with one channel dominance, the coupling of a bound state to its constituents is given in terms of the binding energy by the compositness condition of Weinberg \cite{weinberg}-\cite{danibreak}. 
In the present case, the $DK$ is the dominant channel, but other channels also matter. This is why some changes, although small, were found in the couplings while demanding the same binding energy.

\section{The selfenergy of the $D$ meson}
\subsection{$s$-wave selfenergy}

We shall use the $T=0$ results from the work of \cite{Tolos:2007vh}. There, the
$D$ meson selfenergy is obtained from a selfconsistent coupled channel
calculation, whose driving term is a broken SU(4) $s$-wave Weinberg Tomozawa
interaction supplemented by an attractive isoscalar-scalar term. The introduction of a supplementary scalar-isoscalar interaction in the diagonal $DN$ channel, the $\Sigma_{DN}$ term, which is prevalent in the QCD sum rule and mean-field approaches, is, however, a subject of controversy. Its effect on the $D$ meson self-energy was studied in \cite{mizutani} and \cite{Tolos:2007vh}, and compared to the case where this term was neglected. It was found in \cite{Tolos:2007vh} that the results obtained with or without this term of uncertain origin were qualitatively identical, such that the phenomenology did not allow to draw any conclusion concerning it. The differences found there between the two options are far smaller than the uncertainties that we have in the present problem from other sources and, hence, we take the option of ignoring this term. The Bethe Salpeter
equation is then solved using a cut off regularization, which is fixed by
reproducing the position and width of the $\Lambda_c(2593)$ resonance. As a
result, a new resonance in the $I=1$ channel is generated around
2800 MeV, the $\Sigma_c(2800)$. The coupled channel structure includes: $\pi\Lambda_c$, $\pi\Sigma_c$,
$DN$, $\eta \Lambda_c$, $K\Xi_c$, $\eta\Sigma_c$, $K\Xi'_c$, $D_s\Lambda$, 
$D_s\Sigma$, $\eta'\Lambda_c$ and $\eta'\Sigma_c$.   

The in medium $s$-wave $DN$ amplitude accounts for Pauli blocking effects on the
nucleons in the $DN$ channel, mean-field bindings of baryons via a
$\sigma$-$\omega$ model, and renormalization of $\pi$ and the $D$ through their
corresponding selfenergies in the intermediate propagators. The $s$-wave $D$ selfenergy
is obtained iteratively following a selfconsistent procedure as one integrates the
in medium $s$-wave $DN$ amplitude over the nucleon Fermi sea $n(p)$:
\begin{eqnarray}
\Pi_{D}^{(s)}(q^0,{\vec q},\rho)= \int \frac{d^3p}{(2\pi)^3} \,n(p)\,
[{\tilde T}^{(I=0)}_{DN}(P^0,\vec{P},\rho) +
3{\tilde T}^{(I=1)}_{DN}(P^0,\vec{P},\rho)]\ , \label{eq:selfd}
\end{eqnarray}
where $\tilde T^{(I=0,1)}$ stands for the in medium $s$-wave $DN$ amplitudes in $I=0$ and $I=1$. The quantities $P^0=q^0+E_N(\vec{p})$ and $\vec{P}=\vec{q}+\vec{p}$ are
the total energy and momentum of the $D N$ pair in the nuclear
matter rest frame, with $E_N(\vec{p}\,)$ being the single-particle nucleon
energy and the values ($q^0$, $\vec{q}\,$) 
the energy and momentum of the $D$ meson also in this
frame. For more details see \cite{Tolos:2007vh}.

In fact, the $s$-wave $D$ meson selfenergy in the medium was initially studied in
\cite{Tolos:2004yg,tolos,lutzmed} and further work was done in \cite{mizutani,Tolos:2007vh}. There are novelties in the approach of \cite{mizutani,Tolos:2007vh} with respect to the earlier works. Indeed, for consistency with the reduction from $t$-channel vector-meson exchanges to a zero-range Weinberg-Tomozawa form, corroborated with explicit cancellations of terms, the model of \cite{mizutani,Tolos:2007vh} removes a factor $k^\mu k^\nu/M_V^2$, which was used in \cite{lutzmed}, and by means of which a better width for the $\Lambda_c(2593)$ is obtained in \cite{mizutani,Tolos:2007vh}. Also in \cite{mizutani,Tolos:2007vh} the authors use a conventional momentum cut-off regularization that was found to be more appropiate than the dimensional method in view of its application to meson-baryon scattering in the nuclear medium where Lorentz covariance is manifestly broken, as discussed in Section 2.

With respect to the work of \cite{Tolos:2004yg,tolos} there are also novelties in \cite{mizutani,Tolos:2007vh}. In the exploratory work of \cite{Tolos:2004yg,tolos}, the free amplitudes were constructed from separable coupled channel interactions obtained from chiral motivated Lagrangians upon replacing the $s$ quark by the $c$ quark. While these works give the first indication that the $\Lambda_c(2593)$ could have a dynamical origin, they ignored the strangeness degree of freedom due its very construction. Therefore, the $\pi$ and $K$ (Goldstone) mesons were not treated on an equal footing, and the role of some channels that would appear in the corresponding SU(4) meson and baryon multiplets was ignored.

\subsection{$p$-wave selfenergy}

 We start by recalling the $SU(3)$ chiral
Lagrangian \cite{ecker, ulfreport} for the coupling of pseudoscalar mesons of
the octet of the $\pi$ to the baryon octet of the proton $p$
\begin{equation}
{\cal L}_1^{(B)}=\frac{1}{2}D<\bar{B} \gamma^\mu \gamma_5\{u_\mu,B\}>+\frac{1}{2}F<\bar{B} \gamma^\mu \gamma_5[u_\mu,B]> \ ,
\label{lagranBP}
\end{equation}
where
\begin{eqnarray}
u^2=U=e^{i\frac{\sqrt{2}\phi}{f}} \ ,
\end{eqnarray}
with $\phi$ the usual $SU(3)$ matrix of the meson fields, $f=1.15 f_{\pi}$ with $f_{\pi}=93~{\rm MeV}$  and
\begin{eqnarray}
u_\mu=i u^\dagger \partial_\mu U u^\dagger= -\frac{\sqrt{2}}{f} \partial_\mu \phi +O(\phi^3) \ .
\end{eqnarray}

The $B$ and $\bar{B}$ terms stand for the $SU(3)$ matrices of the baryon fields and $< >$ 
for the trace in $SU(3)$. Hence, at the one meson field level we have
\begin{eqnarray}
{\cal L}_1^{(B)}=&-\frac{1}{\sqrt{2}f}D<\bar{B}\gamma^\mu \gamma_5 \{\partial_\mu \phi, B\}>-\frac{1}{\sqrt{2}f}F<\bar{B}\gamma^\mu \gamma_5 [\partial_\mu \phi, B]> \ .
\end{eqnarray}

\begin{figure}[htb]
     \centering
      \epsfxsize = 6cm
      \epsfbox{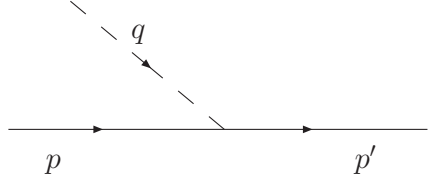}
      \caption{\small Meson-baryon scattering with an outgoing baryon. The labels $p$, $p'$ and $q$ refer to the momenta of the initial baryon, final baryon and meson respectively.}
        \label{fig1}
\end{figure}

The $\bar{u}(\vec{p}\,')\gamma^\mu \gamma_5 u(\vec{p})$ vertex, assuming 
$\vec{p}\simeq 0$ since it will be the momentum of a nucleon in the Fermi sea, 
can be expressed up to $O(1/M^2)$ in terms of the $\vec{\sigma}$ operator such that 
the $T$ matrix corresponding to the diagram of Fig.~\ref{fig1}
is given by
\begin{eqnarray}
-it=\frac{1}{ \sqrt{2}f}\vec{\sigma}\cdot\vec{q}\left(1-\frac{q^0}{2M'}\right)[(D+F)<\bar{B}\phi B>+(D-F)<\bar{B}B\phi>] \ ,
\end{eqnarray}
with $M'$ the mass of the outgoing baryon in Fig.~\ref{fig1}.
We take $D=0.80$ and $F=0.46$ from \cite{ulfepj,close,borasoy}. In order 
to evaluate the  coupling of the $D$ meson to the nucleon and $\Lambda_c$, $\Sigma_c$ we 
use $SU(4)$ symmetry.  We couple the $20$-plet of the baryons, to which the 
nucleon belongs, to the $\bar{20}$ representation of the antibaryons in order to give the 
$15$-plet of the mesons of the $\pi$ and the $D$ \cite{PDG}. By using the $SU(4)$ 
Clebsch-Gordan coefficients of \cite{su4}, we have two independent irreducible 
matrix elements which can be related to the $D$ and $F$ coefficients. The result
 is that the couplings $D^0 p\to \Lambda^+_c$, $D^0 p\to \Sigma^+_c$, 
 $D^+ p\to \Sigma^{++}_c$, $D^0 n\to \Sigma^0_c$, $D^+ n\to \Lambda^+_c$, 
 $D^+ n\to \Sigma^+_c$ are identical to those of $K^- p\to \Lambda$, 
 $K^- p\to\Sigma^0$, $\bar{K}^0 p\to \Sigma^{+}$, $K^- n\to \Sigma^-$, 
 $\bar{K}^0 n\to \Lambda$, $\bar{K}^0 n\to \Sigma^0$   given in 
 \cite{angelsfi} by
\begin{equation}
-iV_{DNY}=\vec{\sigma}\cdot\vec{q} \left( 1-\frac{q^0}{2M'} \right) \left[\alpha\frac{D+F}{2f}+\beta\frac{D-F}{2f}\right] \ ,
\end{equation}
with the coefficients $\alpha$, $\beta$ of the Table \ref{coefficients}.  
\begin{table}[ht]
\centering
\vspace{0.5cm}
\begin{tabular}{c|cccccc}
        & $D^0p\to\Lambda^+_c$ & $D^0p\to \Sigma^+_c$ & $D^0 n\to \Sigma^0_c$
&
$D^+ n \to \Lambda^+_c$ & $D^+ n \to \Sigma^+_c$ &
$D^+ p \to \Sigma^{++}_c$ \\
        \hline
	\hline
$\alpha$ & $-\frac{2}{\sqrt{3}}$ & 0 & 0 & $-\frac{2}{\sqrt{3}}$ & 0
&
0 \\
$\beta$ & $\frac{1}{\sqrt{3}}$ & 1 & $\sqrt{2}$ &
$\frac{1}{\sqrt{3}}$
& $-1$ & $\sqrt{2}$ \\
\hline
\end{tabular}
\caption{\small Coefficients for the $DNY$ couplings}
\label{coefficients}
\end{table}

We also take into account the coupling of the $D$ meson with $\Sigma_c^*(2520)$ and $N$, in analogy to the $p$-wave interaction of pions and kaons with nucleons. For pions and kaons it was shown that the
$N^{-1}\Delta$ and $N^{-1}\Sigma^*(1385)$ excitations, respectively, are relevant for the calculation of the $p$-wave self-energy. Once again, we obtain the same result as in \cite{angelsfi} for the 
$N^{-1}\Sigma^*(1385)$ 
\begin{equation}
-iV_{DNY^*}=a\vec{S}^\dagger\cdot \vec{q} \left(\frac{2\sqrt{6}}{5} \frac{D+F}{2f}\right) \ ,
\label{VDNY}
\end{equation}
with $\vec{S}^\dagger$ being the spin $1/2\to 3/2$ transition operator and $a$ the 
coefficients given in Table \ref{couplingsDNY*}.
\begin{table}[ht]
\centering
\vspace{0.5cm}
\begin{tabular}{c|cccc}
        & $D^0p\to\Sigma^{*+}_c$ & $D^0 n\to \Sigma^{* 0}_c$ & 
$D^+ p \to \Sigma^{* ++}_c$ & $D^+ n \to \Sigma^{* +}_c$ \\
        \hline
	\hline
$a$ & $-\frac{1}{\sqrt{2}}$ & $-1$ & $-1$ & $\frac{1}{\sqrt{2}}$ \\
\hline
\end{tabular}
\caption{\small Coefficient for the $DN\Sigma^*_c(2520)$ couplings }
\label{couplingsDNY*}
\end{table}

\begin{figure}[htb]
     \centering
      \epsfxsize = 4cm
      \epsfbox{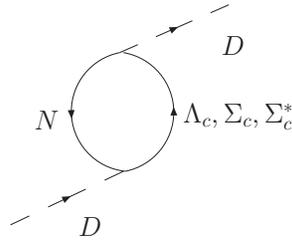}
      \caption{\small $p$-wave selfenergy diagram of the $D$ meson.}
        \label{fig2}
\end{figure}

 With all those couplings, we can readily evaluate the $p$-wave $D$ selfenergy given by 
the diagram of the Fig.~\ref{fig2}, in complete analogy to \cite{angelsfi}. The $p$-wave contribution coming from the $N^{-1}\Lambda_c$ and $N^{-1}\Sigma_c$ excitations reads
\begin{eqnarray}
\Pi^{(p)}_{D^0}(q^0,\vec{q},\rho)& =&\lbrace
\frac{1}{2}B^2_{D^0 p \Lambda^+_c} {\vec q}\,^2
U_{\Lambda^+_c}(q^0,\vec{q},\rho) 
+ \frac{1}{2} B^2_{D^0 p \Sigma^+_c}
  {\vec q}\,^2 U_{\Sigma^+_c}(q^0,\vec{q},\rho) \nonumber\\
&+&  \frac{1}{2} B^2_{D^0 n \Sigma^{0}_c}
 {\vec q}\,^2 U_{\Sigma^0_c}(q^0,\vec{q},\rho)\rbrace F^2_L(q^2) \ ,
\label{selfen}
\end{eqnarray}
where 
\begin{equation}
B_{DNY}=\left(1-\frac{q^0}{2 M_Y}\right)\left[\alpha \frac{D+F}{2f}+\beta\frac{D-F}{2f}\right] \ ,
\end{equation}
and $U$ is the Lindhard function for the $N^{-1}Y$ excitation given by
\begin{eqnarray}
{\rm Re}\, U_Y(q^0,\vec{q},\rho) &=&
\frac{3}{2}\rho \frac{M_Y}{q p_F} \left\{ z + \frac{1}{2}(1-
z^2) \ln \frac{\mid z + 1\mid}{\mid z - 1 \mid} \right\}
\nonumber \\
{\rm Im}\, U_Y(q^0,\vec{q},\rho) &=& -\pi \frac{3}{4} \rho
\frac{M_Y}{q p_F} \left\{ (1-z^2) \theta(1-\mid z \mid)
\right\} \\
z &=&\left( q^0 - \frac{q^2}{2M_Y} - (M_Y-M) \right) \frac{M_Y}{q
p_F} \nonumber \ ,
\end{eqnarray}
with $\rho=\rho_n+\rho_p$, the nuclear density, 
$p_F=(3\pi^2 \rho/2)^{1/3}$ the Fermi momentum, $M_Y$ the hyperon mass 
and $M$ the nucleon mass. The same result holds for the $p$-wave $D^+$ selfenergy 
ignoring small mass differences between particles of the same isospin 
multiplet.

The $p$-wave selfenergy due to the excitation of the 
decuplet is also readily evaluated and we find
\begin{equation}
\Pi^{(p)*}_{D^0}(q^0,\vec{q},\rho)=\lbrace
\frac{1}{3}C^2_{D^0 p \Sigma^{+*}_c} {\vec q}\,^2
U_{\Sigma^{+*}_c}(q^0,\vec{q},\rho) 
+ \frac{1}{3} C^2_{D^0 n \Sigma^{0 *}_c}
  {\vec q}\,^2 U_{\Sigma^{0 *}_c}(q^0,\vec{q},\rho) \rbrace F^2_L(q^2)\ ,
  \label{selp}
\end{equation}
where 
\begin{equation}
C_{DNY}=a f_Y^*\frac{2\sqrt{6}}{5}\frac{D+F}{2f} \ ,
\end{equation}
with $a$ given in Table \ref{couplingsDNY*} and $f_Y^*$ being a recoil factor 
\cite{angelsfi}, which we approximate by $f_Y^*\simeq (1-M_D/M_Y)$. 

In Eqs. (\ref{selfen}) and (\ref{selp}), we include a form factor of monopole type at the $D$ meson-baryon vertices by analogy to the one accompanying the Yukawa $\pi NN$ vertex  \cite{angelsfi}-\cite{Toki}.
\begin{eqnarray}
F_L(q^2)=\frac{\Lambda^2}{\Lambda^2+\vec{q}\,^2}\hspace{0.5cm}\mathrm{with}\hspace{0.2cm}\Lambda=1.05~{\rm GeV}.
\label{eq:light}
\end{eqnarray}

This form factor is suited for light hadrons, i.e., pion excitation of $ph$. However, it is unlikely that the range of $\Lambda$ is the same when dealing with $D$ mesons. There are indications that the form factor to account for off shell $D$ mesons requires a value of $\Lambda$ substantially larger \cite{Navarra}. We shall come back to this point at the end of 
the Results Section, reevaluating results with the heavy meson form factor and analyzing the uncertainties.

With regards to the $p$-wave $D^+$ selfenergy,  it turns out to be  the 
same as for $D^0$ in symmetric nuclear matter $\rho_n=\rho_p$.

\begin{figure}[htb]
     \centering
      \epsfxsize = 15cm
      \epsfbox{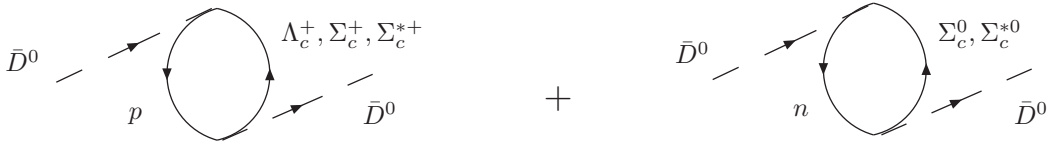}
      \caption{\small $p$-wave selfenergy diagrams of the $\bar{D}$ meson.}
        \label{fig3}
\end{figure}

  For the $\bar D$ meson, we note that the $p$-wave $\bar{D}$ 
selfenergy would correspond to the diagrams in Fig.~\ref{fig3},
which involve the difference between the sum of $\bar{D}$ and $Y$ masses, and 
the nucleon mass. The contribution of those diagrams is negligible due to the large mass of the $\bar{D}$ and $\Sigma_c$, 
$\Lambda_c$. The same holds for the $p$-wave $\bar{D}$ 
selfenergy coming from the $N^{-1}\Sigma_c^*(2520)$ excitation.

\subsection{ The $D$ meson spectral function} 

The selfenergy of a $D$ meson in nuclear matter is given by the coherent sum of the $s$-wave and $p$-wave selfenergies:
\begin{equation}
\Pi_{D}(q^0,\vec{q},\rho)=\Pi_{D}^{(s)}(q^0,\vec{q},\rho)+\Pi_{D}^{(p)}(q^0,\vec{q},\rho)+\Pi_{D}^{(p)*}(q^0,\vec{q},\rho) \ .
\end{equation}
  Then, the $D$ 
propagator is written in the medium as
\begin{equation}
D_{D}(q^0,\vec{q},\rho)=\frac{1}{{(q^0)}^2-\vec{q}\,^2-m^2_{D}-\Pi_{D}(q^0,\vec{q},\rho)} \ .
\end{equation}
For later purposes, it is convenient to write the $D$ propagator in terms of its 
Lehmann representation \cite{waas,cabrera}
\begin{equation}
D_{D}(q^0,\vec{q},\rho)=\int^\infty_0 d\omega \left\{
\frac{S_{D}(\omega,\vec{q},\rho)}{q^0-\omega+i\eta}-\frac{S_{\bar{D}}(\omega,\vec{q},\rho)}{q^0+\omega-i\eta}\right\} \ ,
\label{propmedium}
\end{equation}
where $S_D$ and $S_{\bar{D}}$ are the spectral functions of $D$ and $\bar{D}$, respectively,
\begin{equation}
S_{D(\bar{D})}(q^0,\vec{q},\rho)=-\frac{1}{\pi}\frac{ {\rm Im} \Pi_{D (\bar{D})} (q^0,\vec{q},\rho) }
{{\vert {(q^0)}^2 -{\vec q\,}^2-m^2_D-\Pi_{D (\bar{D})}(q^0,\vec{q},\rho)\vert}^2} \ .
\end{equation}

 In the calculations we will ignore the selfenergy of the $\bar{D}$. As in 
the case of the $K$ with
respect to the $\bar{K}$, the $\bar{D}$ selfenergy is smaller than that of the
$D$ meson and, more importantly, it has no imaginary part from inelastic channels.
Hence, it does not lead to modifications of the width of the states that we
study here, which is the most striking change that we find. We will discuss this in more detail in what follows.

\begin{figure}[htb]
     \centering
      \epsfxsize = 6cm
      \epsfbox{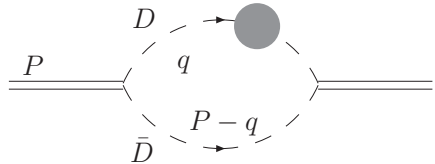}
      \caption{(Color online) \small The $D\bar{D}$ loop function of the scalar meson. The shaded circle indicates the $D$ selfenergy insertion.}
        \label{fig4}
\end{figure}

\begin{figure}
\centering
\includegraphics[width=0.7\textwidth]{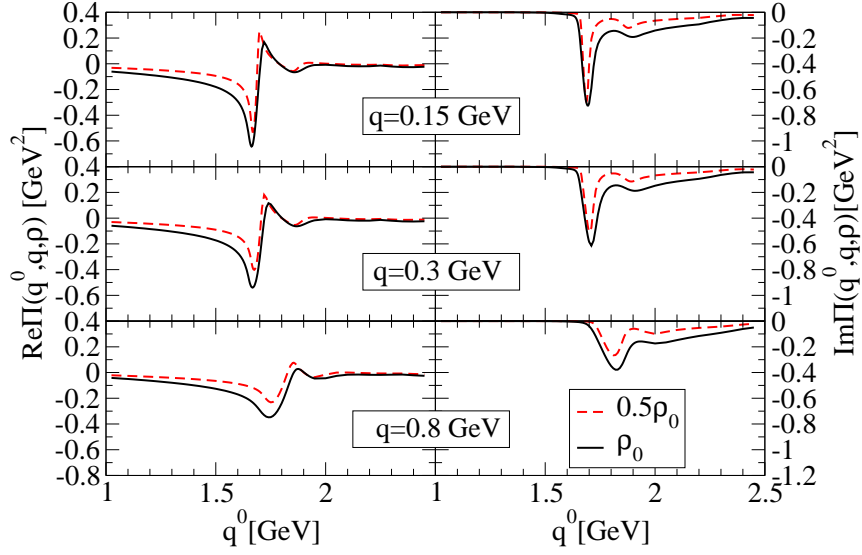}  
\caption{(Color online) \small Real (left column) and Imaginary (right column) part of the $s$-wave $D$ selfenergy for $D$ three-momenta $q=0.15,0.3$ and $0.8~{\rm GeV}$ as a function of the $D$ energy $q^0$ at densities $\rho=0.5\rho_0$ and $\rho=\rho_0$, with $\rho_0=0.17 {\rm fm}^{-3} $ the normal nuclear matter density.}
\label{sself}
\end{figure}

\begin{figure}
\centering
\includegraphics[width=0.7\textwidth]{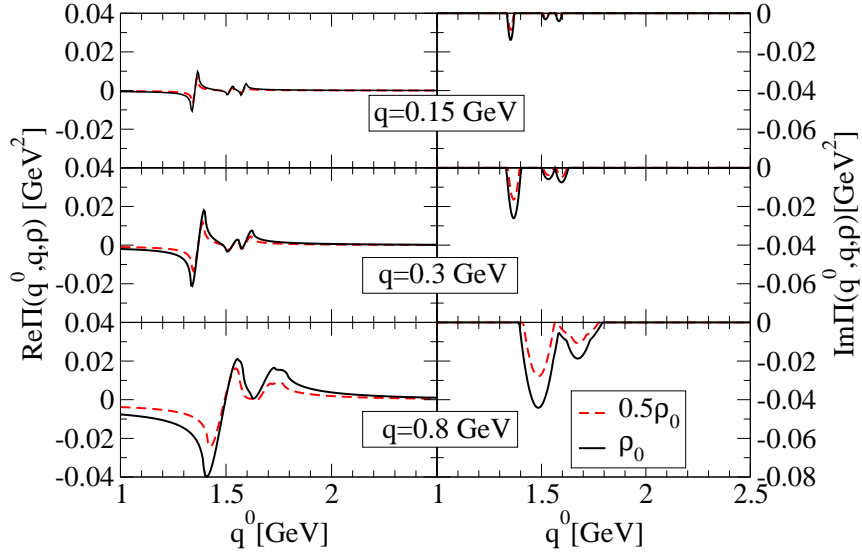}  
\caption{(Color online) \small Real (left column) and Imaginary (right column) part of the $p$-wave $D$ selfenergy for $D$ three-momenta $q=0.15,0.3$ and $0.8~{\rm GeV}$ as a function of the $D$ energy $q^0$ at densities $\rho=0.5\rho_0$ and $\rho=\rho_0$, with $\rho_0=0.17 {\rm fm}^{-3} $ the normal nuclear matter density.}
\label{pself}
\end{figure}

\section{Two meson loop function in the medium}
\label{two}

 The medium modifications are introduced in the two loop meson function by using the dressed two meson propagator in nuclear matter. As an example, let us evaluate the $G$ function in the medium for the $D\bar{D}$ intermediate state:
\begin{equation}
\widetilde{G}(P^0,\vec{P},\rho)=i\int \frac{d^4 q}{(2\pi)^4} D_D (q,\rho) D_{\bar{D}}(P-q,\rho) \ .
\label{loopmedium}
\end{equation}
We shall dress the $D$ propagator and leave the $\bar{D}$ propagator free. 
The reason to neglect the $\bar{D}$ selfenergy in the medium is that it is very small compared to its mass \cite{Tolos:2007vh}. Indeed, the $p$-wave part is negligible as discussed at the end of Subsection 4.2. The $s$-wave part is equally small, but more importantly there is no absorption of $\bar{D}$ by nucleons, meaning that the $\bar{D}N$ does not decay to baryonic resonances which have $c$ quarks instead of $\bar{c}$ . The analogy is clear with the $K$ and $\bar{K}$, where the $\bar{K}$ (analogous to $D$) can undergo absortion reactions $\bar{K}N\to \pi \Lambda$, $\pi \Sigma$, while the $K^+$ (analogous to $\bar{D}$) cannot be absorbed. Altogether this justifies 
to use the free propagator for $\bar{D}$. We must evaluate the loop function
 of the diagram of Fig.~\ref{fig4},
where the blob in the $D$ propagator symbolizes the $D$ selfenergy insertion, 
indicating that we must use the $D$ propagator in the medium. Then, we use 
Eq.~(\ref{propmedium}) for this propagator. Given the large mass of the $D$ 
mesons, we can also neglect the $S_{\bar{D}}$ part in the propagator of 
Eq.~(\ref{propmedium}), which upon $q^0$ integration in the Eq.~(\ref{loopmedium}) 
leads to a contribution of order of $1/(P^0+2\omega_D (q))$, very small compared 
with the part coming from $S_D$. Thus, we can write
\begin{eqnarray}
\widetilde{G}(P^0,\vec{P},\rho)&=&i\int \frac{d^4 q}{(2 \pi)^4} \int^\infty_0 d\omega \frac{S_D(\omega, \vec{q},\rho)}{q^0-\omega+i\eta}
~\frac{1}{(P^0-q^0)^2-{\vec{q}\,}^2-m^2_{\bar{D}}+i\eta}
\nonumber\\
&=& \int \frac{d^3 q}{(2\pi)^3}\int^\infty_0 d\omega \frac{S_D(\omega,\vec{q},\rho)}{P^0-\omega-\omega_{\bar{D}}(\vec{q}\,)+i\eta}
~ \frac{1}{2\omega_{\bar{D}}(\vec{q}\,)} \ ,
\label{propespmedium}
\end{eqnarray}
where $\omega_{\bar{D}}(\vec{q}\,)=({\vec{q}\,}^2+m^2_{\bar{D}})^{1/2}$. We evaluate 
Eq.~(\ref{propespmedium}) with a three momentum cut off of $q_{max}=0.85$ and $0.9$ GeV for the $X(3700)$ and $D_{s0}(2317)$, respectively, 
equivalent to the use of dimensional regularization of \cite{Gamermann:2006nm} 
with the chosen substraction constants.

Following the discussion in the former section, the $D\bar{D}$ loop function 
will appear in the case of the $X(3700)$ hidden charm state. In the case of the 
$D_{s0}(2317)$, we have $K$ instead of $\bar{D}$ and, again, we use its free 
propagator neglecting the small structureless $K$ selfenergy. 

\section{Results}
 
 In Figs. \ref{sself} and \ref{pself} we show the results of the $s$-wave and
$p$-wave selfenergies of the $D$ meson. We perform the calculations first by using the light meson version of the $Dph$ form factor, Eq. (\ref{eq:light}). We will show results with a heavy meson form factor at the end of this section. In Fig.~\ref{sself} the structures around energies 1.7 GeV and 2 GeV correspond to the excitation of the $h \Lambda_c(2593)$ and $h \Sigma_c(2800)$, where the $ \Lambda_c(2593)$ and $ \Sigma_c(2800)$ are $1/2^{*-} $ dynamically generated resonances of the theory and $h$ stands for hole of nucleon. With regard to the $p$-wave contribution, the structures seen in the $p$-wave
selfenergy, with peaks for the imaginary part and oscillations in the real one
around 1.4 to 1.8 GeV,  correspond to the excitation of the different $h Y $  components, with $Y=\Lambda_c$, $\Sigma_c$, $\Sigma^*_c$.

Note that the $p$-wave selfenergy is much smaller than the $s$-wave selfenergy,
even up to momentum as large as $800~{\rm MeV/c}$. The main reason is that
the $s$-wave selfenergy of the mesons goes roughly as the meson mass, from the
Weinberg Tomozawa interaction, while the $p$-wave scales differently, roughly
like the baryon mass. In the case of pions, the $p$-wave selfenergy was more 
important than the $s$-wave \cite{Nieves:1993ev}, but in the case of kaons the
relative importance of the $p$-wave was already smaller \cite{laurap,Ramos:1999ku}.

\begin{figure}[htb]
     \centering
	\includegraphics[width=0.6\textwidth,height=0.4\textwidth]{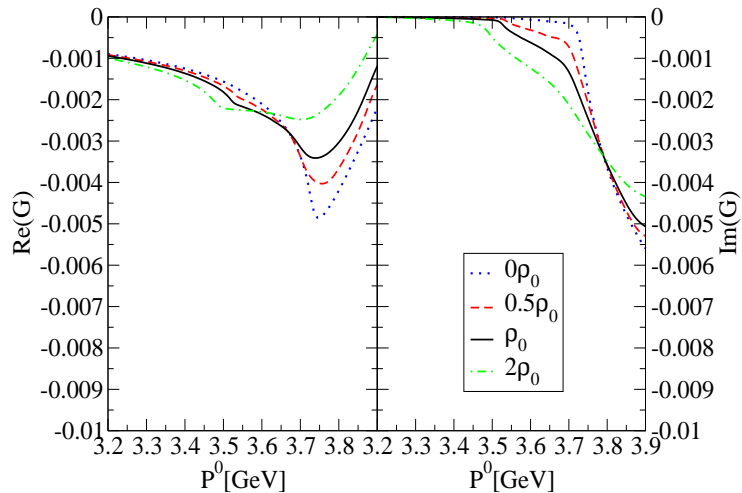}  
      \caption{(Color online) \small Loop function in the medium: ${\rm Re}\,\widetilde{G}(P^0,\vec{P},\rho)$ (left) and
      ${\rm Im}\,\widetilde{G}(P^0,\vec{P},\rho)$ (right) for  $D\bar{D}$, the channel with the largest coupling to the X(3700) meson. $\widetilde{G}(P^0,\vec{P},\rho)$ is given from Eq. (\ref{propespmedium}).}
        \label{fig5}
\end{figure}

\begin{figure}[htb]
     \centering
	\includegraphics[width=0.6\textwidth,height=0.4\textwidth]{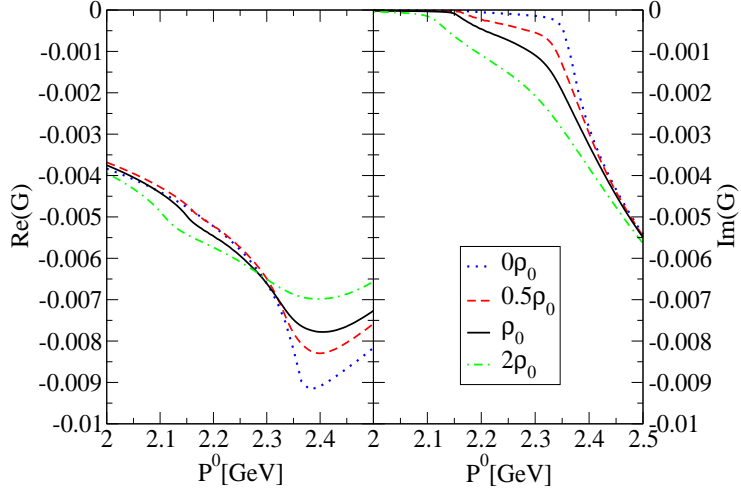}  
      \caption{(Color online) \small Loop function in the medium: ${\rm Re}\,\widetilde{G}(P^0,\vec{P},\rho)$ (left) and
      ${\rm Im}\,\widetilde{G}(P^0,\vec{P},\rho)$ (right) for  $DK$, the channel with the largest coupling to the $D_{s0}(2317)$ meson. $\widetilde{G}(P^0,\vec{P},\rho)$ is given from Eq. (\ref{propespmedium}).}
        \label{fig6}
\end{figure}

\begin{figure}[htb]
     \centering
      \epsfxsize = 8cm
      \epsfbox{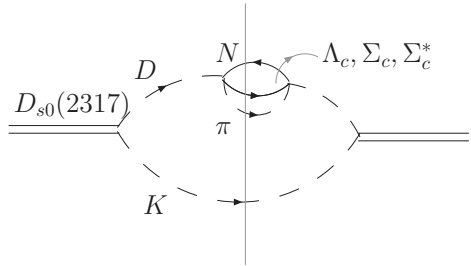}
      \caption{\small Decay channel of the $D_{s0}(2317)$ in the nucleus into $K\pi\Lambda_c$ or $K\pi\Sigma_c$.}
        \label{fig10}
\end{figure}

In Figs. \ref{fig5} and \ref{fig6}, 
 we plot the $\widetilde{G}(P^0,\vec{P},\rho)$ function for $\vec{P}=0$ for the $D\bar{D}$ and $DK$ loops for different densities. The effects of the density in the loop function are clearly visible in all cases. The imaginary part is largely increased at lower energies and collects strength below threshold of the $D\bar{D}$ or $DK$ channels, respectively, due to the opening of new many body decay channels of the meson-meson system. As an example, let us take the $DK$ loop function (see Fig. \ref{fig10}). The $D$ is renormalized and accounts for $DN \to \pi\Lambda_c$, $\pi\Sigma_c$ or $DN\to\Lambda_c$, $\Sigma_c$. Hence, the $DK$ loop in the medium accounts for intermediate channels $h\pi\Lambda_c K$, $h\pi\Sigma_c K$ or $h\Lambda_c K$, $h\Sigma_c K$ which have a smaller mass than the $DK$ system and open up at lower energies than $DK$ threshold. The real parts of $\widetilde{G}$ are also sizeably modified around the thresholds as one can see in the figures.
 

\begin{figure}
\centering
\begin{tabular}{cc} 
\includegraphics[width=0.5\textwidth,height=0.4\textwidth]{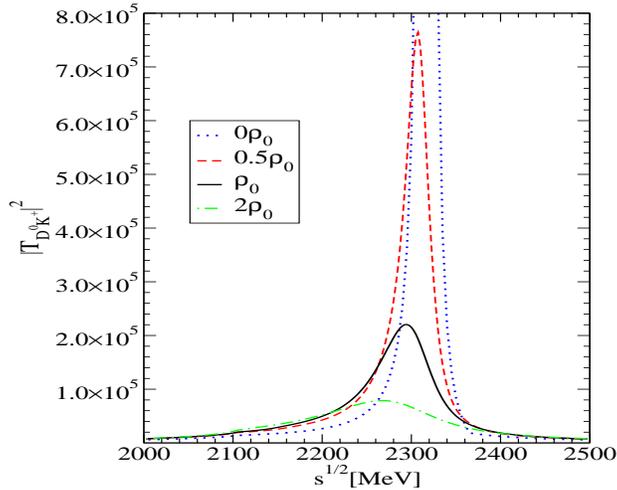}  
\end{tabular}
\caption{(Color online) \small $D_{s0}(2317)$ resonance: $\vert T\vert^2$ for the $D^0K^+ \to D^0K^+$ amplitude for different densities.}
\label{fig9}
\end{figure}

In Fig.~\ref{fig9} we show the ${\vert T \vert}^2$ for the $D^0 K^+ \to D^0 K^+
$  amplitude around the region of $2300~{\rm MeV}$ for different densities. We
can see  that originally, at $\rho=0$, the amplitude exhibits the pole of the 
$D_{s0}(2317)$, with zero width.  This is the reason why it is out of the y-scale 
 in the plot since ${\vert T \vert}^2$  goes up to infinite. As the density increases, we can see a 
slight shift of the mass, of the order of $15~{\rm MeV}$ attraction at
$\rho=\rho_0$.  However, the increase in the width is more spectacular,
which goes  from zero in the free case to about $100~{\rm MeV}$ at
$\rho=\rho_0$, and $200~{\rm MeV}$ at  $\rho=2\rho_0$. This is certainly a
drastic relative effect, and even big in  absolute value. The origin of
the width in the medium is due to the opening of  new channels $DN\to
\Lambda_c, \Sigma_c$ from the $p$-wave selfenergy and   $DN\to
\pi\Lambda_c, \pi\Sigma_c$ from the $s$-wave selfenergy.  On the other  hand, the use of selfconsistency
in the evaluation of the $D$ selfenergy  \cite{mizutani,Tolos:2007vh} generates
also some two nucleon induced $D$ absorption channels like $DNN\to
N\Lambda_c$,  $\pi N\Lambda_c$, $\pi N\Sigma_c$, etc. All these channels
collaborate to make  the $D$ disappear inside the nuclear medium through $DN$
or $DNN$ inelastic  reactions, where the $D$ gets absorbed. 

The study of such decay channels in the nucleus would offer information on 
the coupling of the $D_{s0}(2317)$ to the $DK$, the basic building block of 
the resonance according to the underlying theory that we are using. As an 
example, the exploration of the decay channel of the $D_{s0}(2317)$ in the 
nucleus into $K\pi\Lambda_c$ or $K\pi\Sigma_c$ channels, which corresponds to the cut in the 
diagram of Fig.~\ref{fig10}, 
would provide combined information on the $D_{s0}(2317)\to DK$ coupling and 
the $DN\to \pi\Lambda_c(\Sigma_c)$ cross section.

\begin{figure}
\centering
\begin{tabular}{cc} 
\includegraphics[width=0.5\textwidth, height=0.4\textwidth]{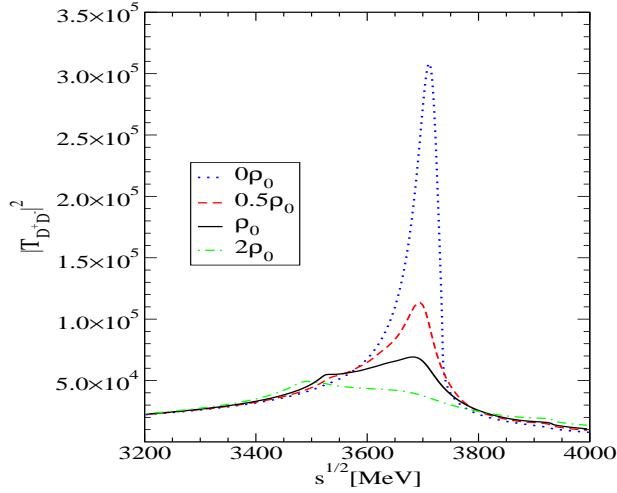}  
\end{tabular}
\caption{(Color online) \small $X(3700)$ resonance: $\vert T\vert^2$ for the $D^0\bar{D}^0 \to D^0\bar{D}^0$ amplitude for different densities.}
\label{fig11}
\end{figure}

The results with the large width for the $D_{s0}(2317)$ are a consequence of the large coupling of this resonance to $DK$. This large coupling is guaranteed by the "compositness condition" of Weinberg \cite{weinberg}-\cite{danibreak}, as far as the resonance is dynamically generated and $DK$ is the main building block. Should this resonance be a $q\bar{q}$ state as suggested in \cite{Lakhina:2006fy} or have any other structure, like $q\bar{q}$ with a mixture of $DK$ components as suggested in \cite{vanBeveren:2005ha,vanBeveren:2005pk}, such large coupling would not appear \cite{weinberg} and, thus, the width in the medium would be much smaller than we predict here.  Hence, investigating the widths in the medium provides extra information on the nature of these resonances. 

In Fig.~\ref{fig11} we show the same results for the hidden charm $X(3700)$
resonance.  The resonance begins with about $60~{\rm MeV}$ in the real axis ($36$ MeV deduced from the imaginary part of the pole position) from its main
decay into the $\eta\eta$, $\eta\eta'$, and $\eta'\eta'$ channels. The difference with \cite{Gamermann:2006nm}, where the
width was much  smaller, is due to the fact that, here, we use the mixing of
mesons  $\eta$ and $\eta'$ \cite{withzou}, which makes  the $\eta_c$ a pure $c\bar{c}$ state rather than pure
$SU(4)$ state of the  15-plet of mesons. The study of this figure shows
that at $\rho=\rho_0$ the  width has become as large as $250~{\rm MeV}$ and,
due to cusps on the multiparticle  channels that open up, the strength of the
resonance in the nuclear medium  acquires a peculiar shape as the density
increases. Here, the main decay  channels in the nucleus are the same as in the
case of the $D_{s0}(2317)$,  replacing the $K$ by a $\bar{D}$. One should note that there is no relationship of the important decay channels to those that dominate the dynamical generation of the state. For example, the $X(3700)$ couples mainly to $D\bar{D}$ but the $D \bar D$ threshold is above the $X(3700)$ in free space and there is no decay into these components. The analogy can be made with the $f_0(980)$ which couples mostly to $K\bar{K}$ \cite{bes,oller} but decays into $\pi\pi$ in free space, since there is no phase space for $K\bar{K}$ decay. The coupling of the $X(3700)$ to the $\eta\eta$, $\eta\eta'$ etc components is small, reflecting the fact that these channels are relatively unimportant in the structure of the resonance. However, because there is large phase space for these decays, they are mostly responsible for the $X(3700)$ width. 

In the medium things might be different. Indeed, in the presence of nucleons we can have $DN\to\pi\Lambda_c$, $\pi \Sigma_c$ as we mentioned above, and new decay channels for the $X(3700)$ appear, such as $X(3700)N\to \bar{D}\pi\Lambda_c$, $\bar{D}\pi\Sigma_c$ [see Fig. \ref{fig10} (changing $K$ by $\bar{D}$)], which have nearly $400$ MeV phase space available. The strong coupling of the $X(3700)$ to $D\bar{D}$ and the opening of these new decay modes makes now the width in the medium sizeable. We note that the trend of changes of $|T|^2$ with nuclear density is typically observed in resonance properties with temperature and/or density \cite{hiller,Rapp1,Rapp2}.

\begin{figure}
\centering
\begin{tabular}{cc} 
\includegraphics[width=7cm,angle=0]{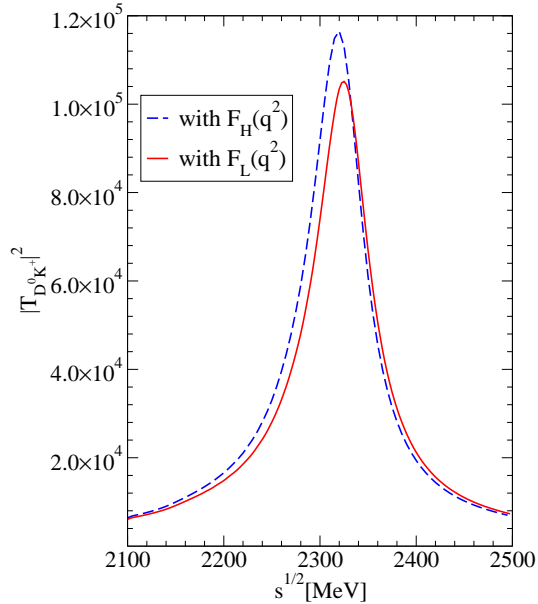}  
\end{tabular}
\caption{(Color online) Comparison of $\vert T \vert^2$ in the case of the $D_{s0}(2317)$ resonance for the two different form factors at $\rho=\rho_0$: type $1$ (dashed line) with $F_H(q^2)$ of Eq. (\ref{eq:heavy}) and type $2$ (solid line) with $F_L(q^2)$ of Eq. (\ref{eq:light}).}
\label{fignew}
\end{figure}

    As mentioned at the end of Subsection 4.2, in order to analyze the uncertainties linked to the use of different form factors, we reevaluate the results using a form factor for the off shell $D$ mesons which we obtain from the work of \cite{Navarra}. The heavy meson form factor is now
   \begin{eqnarray}
   F_H(q^2)=\frac{\Lambda^2_D-m^2_D}{\Lambda^2_D-q^2}\hspace{0.5cm}\mathrm{with}\hspace{0.2cm}\Lambda_D=3.5\,\mathrm{GeV}.
   \label{eq:heavy}
   \end{eqnarray}
The value of $q^2$ in $F_H(q^2)$ is taken for the configurations that give rise to the width in diagram of Fig. \ref{fig10} when the $D$ selfenergy comes from particle-hole ($ph$) excitation of Fig. \ref{fig2}. This means that, for the case of the $D_{s0}(2317)$, we place the $K$ and $\Lambda_c h$, $\Sigma_c h$ or $\Sigma_c^* h$ on shell. This occurs for any value of the running variable $q$ at a value of $q^0$ given by 
\begin{eqnarray}
q^0=M_{D_{s0}}-\omega_K(q_{on})\hspace{0.5cm}\mathrm{with}\hspace{0.2cm}q_{on}=\frac{\lambda^{1/2}((M_{D_{s0}}+m_N)^2,M^2,m_K^2)}{2\, (M_{D_{s0}}+m_N)}\,
\end{eqnarray}
where $M$ can be $M_{\Lambda_c}$,  $M_{\Sigma_c}$ or $M_{\Sigma^*_c}$ depending on the type of $ph$ excitation.
We anticipate that, given the minor relevance of the $p$-wave selfenergy, the differences with respect to the former calculation using the light meson form factor will be small. This is indeed the case as can be seen in Fig. \ref{fignew}, where we show $\vert T \vert^2$ for the case of the $D_{s0}(2317)$ for the two types of form factor. As we can see there are small differences in the position of the peak and no difference in the width.

  As indicated before, in Figs.~\ref{fig9} and ~\ref{fig11} we have shown the 
   squared amplitudes for different densities to facilitate the comparison. As indicated
   at the end of Section $2$, we also evaluate the uncertainties in the results. We show in 
   Tables ~\ref{uncer1} and ~\ref{uncer2} the results obtained for the
   mass and width of the resonances at different densities, with their uncertainties 
   obtained from a Monte Carlo sampling of the parameters $f_{\pi}$ and $f_{D}$ as 
   indicated in Section $2$ and evaluated from the plots of $|T|^2$ in the real axis. The results in these tables are all evaluated using the heavy meson form factor of Eq. (\ref{eq:heavy}). The relevant information from these 
   tables is that the differences between the widths at $\rho=0$ or $\rho=\rho_0$ for both resonances are 
   much bigger than the uncertainties in the width from uncertainties in the model.
   In the case of the masses, we do not see appreciable shift in the mass compared to the uncertainties for 
   the case of the $D_{s0}(2317)$, see Table \ref{uncer1}. For the $X(3700)$, the shift from $\rho=0$ to $\rho=\rho_0$ is 
   of the order of $70$ MeV, see Table \ref{uncer2}, smaller than the width in the medium. Hence we cannot make any strong point concerning a 
   possible shift of the masses.
   
\begin{table}
\centering
\vspace{0.5cm}
\begin{tabular}{c|cc}
\hline
$\rho$ & $\bar M$[MeV] & $\bar \Gamma$[MeV] \\
\hline
\hline
0.0   & 2316$\pm$5 &0 \\
0.5   & 2306$\pm$17 & 58$\pm$10\\
1.0  & 2295$\pm$23 & 115$\pm$25\\
1.5   & 2283$\pm$25 & 150$\pm$25\\
2.0   & 2274$\pm$31 & 190$\pm$30\\
\hline
\end{tabular}
\caption{ Mass and width for the $D_{s0}(2317)$ at different densities with error bands due to the uncertainties of our model.} 
\label{uncer1}
\end{table}
\begin{table}
\centering
\vspace{0.5cm}
\begin{tabular}{c|cc}
\hline
$\rho$ & $\bar M$[MeV] & $\bar \Gamma$[MeV] \\
\hline
\hline
0.0   & 3710$\pm$18 & 60$\pm$ 10 \\
0.5   & 3691$\pm$ 10 & 135$\pm$20\\
1.0  & 3638$\pm$ 15 & 255$\pm$25\\
1.5   & 3599$\pm$15 & 320$\pm$25\\
2.0   & 3565$\pm$29 & 340$\pm$25\\
\hline
\end{tabular}
\caption{ Mass and width for the $X(3700)$ at different densities with error bands due to the uncertainties of our model.} 
\label{uncer2}
\end{table}

In view of future experiments measuring medium modifications of these 
resonances, we can recall the method that has proved most efficient in 
measuring nuclear widths: the transparency ratio. The direct measurement  in experiments of the in medium
increased width  is not easy  because the 
observed decay channels usually come from the resonances 
that have escaped from the nucleus,  so the density at the decay place is zero
or very small
\cite{Kaskulov:2005uw}. Therefore, one should look at the production rate 
as a function of the mass number normalized to a particular nucleus 
(transparency ratio). This magnitude, which measures the survival probability,
is very sensitive to the absorption rate of the resonance inside the nucleus,  i.e.,
the in medium resonance width. This procedure has been succesfully used for the $\phi$ and $\omega$ production in nuclei in \cite{omegawidth,imai}  with 
the help of relatively easy tools of analysis \cite{kaskulov,rocafi}.

\section{Conclusions}
  We have evaluated the selfenergy of  low lying scalar mesons with open 
and hidden charm in a nuclear medium, concretely of the $D_{s0}(2317)$ and 
the theoretical hidden charm state  $X(3700)$. The many body
calculation has been done following the lines of previous studies in the
renormalization of the light scalar mesons in the nuclear medium. 
The medium effects for the  $D_{s0}(2317)$ and $X(3700)$ resonances are spectacular. Those resonances, which have zero and small width in free space,
respectively, develop widths of the order of 100 and 200
MeV at normal nuclear matter density, respectively. The study also allowed us 
to trace back the reactions in the medium, which are responsible for the decay 
width of these mesons and which could be investigated in future reactions at
hadron facilities. The option of looking at transparency ratios was 
also
suggested as a mean to investigate the widths of these mesons in nuclei. It was
also discussed that the experimental study of this width and the medium
reactions contributing to it provide information on the basic features of the
resonance and the selfenergy of the $D$ meson in a nuclear medium. In other words,
the experimental analysis of those properties is a
valuable test of the dynamics of the $D$ meson interaction with nucleons and
nuclei, and the nature of the charm and hidden charm scalar resonances, all of 
them topics which are subject of much debate at present. The results obtained 
here should stimulate experimental work in hadron facilities, in particular at 
FAIR \cite{FAIR}, where the investigation of charm physics is one of the priorities.

\section*{Acknowledgments}
We would like to thank Carmen Garcia Recio for her help with the practice of SU(4) algebra. R. M.  wishes to acknowledge support from the Ministerio de Educacion in the
in the program of FPI. L.T. acknowledges support from the BMBF project
``Hadronisierung des QGP und dynamik von hadronen mit charm quarks''  (ANBest-P
and BNBest-BMBF 98/NKBF98), the ``RFF-Open and hidden
charm at PANDA'' project from the Rosalind Franklin Programme of the
University of Groningen (The Netherlands) and the Helmholtz
International Center for FAIR within the framework of the LOEWE
program by the State of Hesse (Germany).
This work is partly supported by
DGICYT Contract No. BFM2003-00856 and  is part of the EU Integrated Infrastructure
Initiative
Hadron Physics Project under Contract No. RII3-CT-2004-506078.


\begin{thebibliography}{99}
    
\bibitem{Post:2003hu}
  M.~Post, S.~Leupold and U.~Mosel,
  Nucl.\ Phys.\  A {\bf 741} (2004) 81
  
\bibitem{Kaskulov:2005uw}
  M.~Kaskulov and E.~Oset,
  Phys.\ Rev.\  C {\bf 73} (2006) 045213


\bibitem{kaskulov}
  M.~Kaskulov, E.~Hernandez and E.~Oset,
  Eur.\ Phys.\ J.\  A {\bf 31} (2007) 245
  
\bibitem{omegawidth}
  M.~Kotulla {\it et al.}  [CBELSA/TAPS Collaboration],
  Phys.\ Rev.\ Lett.\  {\bf 100} (2008) 192302
  \bibitem{Wolf}
  G.~Wolf, B.~Friman and M.~Soyeur,
  Nucl.\ Phys.\  A {\bf 640} (1998) 129
 
\bibitem{Broniowski}
  W.~Broniowski, W.~Florkowski and B.~Hiller,
  Eur.\ Phys.\ J.\  A {\bf 7} (2000) 287
\bibitem{Broniowski2}
  W.~Broniowski, W.~Florkowski and B.~Hiller,
  Acta Phys.\ Polon.\  B {\bf 30} (1999) 1079
\bibitem{Beveren}
  E.~van Beveren, T.~A.~Rijken, K.~Metzger, C.~Dullemond, G.~Rupp and J.~E.~Ribeiro,
  Z.\ Phys.\  C {\bf 30} (1986) 615;
  \bibitem{Tornqvist}
  ~N.~A.~Tornqvist,
  Z.\ Phys.\  C {\bf 68} (1995) 647

\bibitem{Weinstein}
 ~J.~D.~Weinstein and N.~Isgur,
  Phys.\ Rev.\ Lett.\  {\bf 48} (1982) 659
  \bibitem{Jaffe}
  ~ R.~L.~Jaffe,
  Phys.\ Rev.\  D {\bf 15} (1977) 267
  \bibitem{Black}
  ~ D.~Black, A.~H.~Fariborz, F.~Sannino and J.~Schechter,
  Phys.\ Rev.\  D {\bf 58} (1998) 054012
  \bibitem{Fariborz}
  ~A.~H.~Fariborz,
  Int.\ J.\ Mod.\ Phys.\  A {\bf 19} (2004) 5417
  \bibitem{Fariborz2}
   ~A.~H.~Fariborz, R.~Jora and J.~Schechter,
  Phys.\ Rev.\  D {\bf 72} (2005) 034001;
 
 \bibitem{Giacosa}
  ~F.~Giacosa,
  Phys.\ Rev.\  D {\bf 74} (2006) 014028;
  \bibitem{Volkov}
  ~M.~K.~Volkov and V.~L.~Yudichev,
  Int.\ J.\ Mod.\ Phys.\  A {\bf 14} (1999) 4621
 \bibitem{Black2}
   ~D.~Black, A.~H.~Fariborz and J.~Schechter,
  Phys.\ Rev.\  D {\bf 61} (2000) 074001
   \bibitem{Teshima}
  ~T.~Teshima, I.~Kitamura and N.~Morisita,
  J.\ Phys.\ G {\bf 28} (2002) 1391
  \bibitem{Teshima2}
  ~T.~Teshima, I.~Kitamura and N.~Morisita,
  J.\ Phys.\ G {\bf 30} (2004) 663
  \bibitem{Fariborz3}
  ~A.~H.~Fariborz,
  Int.\ J.\ Mod.\ Phys.\  A {\bf 19} (2004) 2095
  \bibitem{Fariborz4}
  ~A.~H.~Fariborz,
  Phys.\ Rev.\  D {\bf 74} (2006) 054030 
  \bibitem{Giacosa2}
  ~F.~Giacosa, T.~Gutsche, V.~E.~Lyubovitskij and A.~Faessler,
  Phys.\ Lett.\  B {\bf 622} (2005) 277;%
  \bibitem{Napsuciale}
 ~M.~Napsuciale and S.~Rodriguez,
  Phys.\ Rev.\  D {\bf 70} (2004) 094043

\bibitem{npa} J.~A.~Oller and E.~Oset,
  Nucl.\ Phys.\  A {\bf 620} (1997) 438
  [Erratum-ibid.\  A {\bf 652} (1999) 407]
  
\bibitem{iam}
  J.~A.~Oller, E.~Oset and J.~R.~Pelaez,
  Phys.\ Rev.\  D {\bf 59} (1999) 074001
  [Erratum-ibid.\  D {\bf 60} (1999\ ERRAT,D75,099903.2007) 099906]
  ~J.~R.~Pelaez,
  Phys.\ Rev.\ Lett.\  {\bf 92} (2004) 102001

  

\bibitem{kaiser}
  N.~Kaiser,
  Eur.\ Phys.\ J.\  A {\bf 3} (1998) 307
\bibitem{sigmaulf}
  V.~Bernard, U.~G.~Meissner and I.~Zahed,
  Phys.\ Rev.\ Lett.\  {\bf 59} (1987) 966
\bibitem{piulf}
  V.~Bernard, U.~G.~Meissner and I.~Zahed,
  Phys.\ Rev.\  D {\bf 36} (1987) 819
\bibitem{Hatsuda:1999kd}
T.~Hatsuda, T.~Kunihiro and H.~Shimizu,
Phys.\ Rev.\ Lett.\  {\bf 82} (1999) 2840

\bibitem{Jido:2000bw}
D.~Jido, T.~Hatsuda and T.~Kunihiro,
Phys.\ Rev.\ D {\bf 63} (2001) 011901


\bibitem{pipi}
P.~Schuck, W.~Norenberg and G.~Chanfray,
Z.\ Phys.\ A {\bf 330} (1988) 119 ;
~R.~Rapp, J.~W.~Durso and J.~Wambach,
Nucl.\ Phys.\ A {\bf 596} (1996) 436 ;
~Z.~Aouissat, R.~Rapp, G.~Chanfray, P.~Schuck and J.~Wambach,
Nucl.\ Phys.\ A {\bf 581} (1995) 471; 
~H.~C.~Chiang, E.~Oset and M.~J.~Vicente-Vacas,
Nucl.\ Phys.\ A {\bf 644} (1998) 77 ;
~Z.~Aouissat, G.~Chanfray, P.~Schuck and J.~Wambach,
Phys.\ Rev.\ C {\bf 61} (2000) 012202;
~D.~Davesne, Y.~J.~Zhang and G.~Chanfray,
Phys.\ Rev.\ C {\bf 62} (2000) 024604


\bibitem{Cabrera:2005wz}
  D.~Cabrera, E.~Oset and M.~J.~Vicente Vacas,
  Phys.\ Rev.\  C {\bf 72} (2005) 025207
  
\bibitem{pidospi}
F.~Bonutti {\it et al.}  [CHAOS Collaboration],
Phys.\ Rev.\ Lett.\  {\bf 77} (1996) 603;
~F.~Bonutti {\it et al.}  [CHAOS Collaboration],
Nucl.\ Phys.\ A {\bf 638} (1998) 729; 
~P.~Camerini, N.~Grion, R.~Rui and D.~Vetterli,
Nucl.\ Phys.\ A {\bf 552} (1993) 451 
[Erratum-ibid.\ A {\bf 572}  (1993) 791];
~F.~Bonutti {\it et al.}  [CHAOS Collaboration],
Phys.\ Rev.\ C {\bf 60}  (1999) 018201;
~A.~Starostin {\it et al.}  [Crystal Ball Collaboration],
Phys.\ Rev.\ Lett.\  {\bf 85} (2000) 5539


\bibitem{Messchendorp:2002au}
J.~G.~Messchendorp {\it et al.},
Phys.\ Rev.\ Lett.\  {\bf 89} (2002) 222302

\bibitem{VicenteVacas:1999xx}
  M.~J.~Vicente Vacas and E.~Oset,
  Phys.\ Rev.\  C {\bf 60} (1999) 064621

\bibitem{Roca:2002vd}
L.~Roca, E.~Oset and M.~J.~Vicente Vacas,
Phys.\ Lett.\ B {\bf 541} (2002) 77

\bibitem{Muhlich:2004zj}
P.~Muhlich, L.~Alvarez-Ruso, O.~Buss and U.~Mosel,
Phys.\ Lett.\ B {\bf 595} (2004) 216

\bibitem{Oset:2000ev}
  E.~Oset and M.~J.~Vicente Vacas,
  Nucl.\ Phys.\  A {\bf 678} (2000) 424
  
\bibitem{Kolomeitsev:2003ac}
  E.~E.~Kolomeitsev and M.~F.~M.~Lutz,
  Phys.\ Lett.\  B {\bf 582} (2004) 39
  
\bibitem{Hofmann:2003je}
  J.~Hofmann and M.~F.~M.~Lutz,
  Nucl.\ Phys.\  A {\bf 733} (2004) 142
  
\bibitem{Guo:2006fu}
  F.~K.~Guo, P.~N.~Shen, H.~C.~Chiang and R.~G.~Ping,
  Phys.\ Lett.\  B {\bf 641} (2006) 278
  
\bibitem{Gamermann:2006nm}
  D.~Gamermann, E.~Oset, D.~Strottman and M.~J.~Vicente Vacas,
  Phys.\ Rev.\  D {\bf 76} (2007) 074016
  
\bibitem{NievesDs}
  J.~M.~Flynn and J.~Nieves,
  Phys.\ Rev.\  D {\bf 75} (2007) 074024
\bibitem{Abe:2007sy}
  K.~Abe {\it et al.}  [Belle Collaboration],
  arXiv:0708.3812 [hep-ex]
  
  
\bibitem{Gamermann:2007mu}
  D.~Gamermann and E.~Oset,
 Eur. Phys. J. A {\bf 36} (2008) 189   
  
\bibitem{Bando:1984ej}
 M.~Bando, T.~Kugo, S.~Uehara, K.~Yamawaki and T.~Yanagida,
 Phys.\ Rev.\ Lett.\  {\bf 54} (1985) 1215.

\bibitem{Bando:1985rf}
 M.~Bando, T.~Kugo and K.~Yamawaki,
 Nucl.\ Phys.\  B {\bf 259} (1985) 493.
 \bibitem{hidden3}
  M.~Harada and K.~Yamawaki,
  Phys.\ Rept.\  {\bf 381}, 1 (2003)
  
\bibitem{ulfvec}
 U.~G.~Meissner,
 Phys.\ Rept.\  {\bf 161}, 213 (1988).
\bibitem{Walliser:1992vx}
  H.~Walliser,
  Nucl.\ Phys.\  A {\bf 548} (1992) 649.
 
\bibitem{Jido}
 D.~Jido, J.~A.~Oller, E.~Oset, A.~Ramos and U.~G.~Meissner,
 Nucl.\ Phys.\  A {\bf 725} (2003) 181
 
\bibitem{bennhold}
 E.~Oset, A.~Ramos and C.~Bennhold,
 Phys.\ Lett.\  B {\bf 527} (2002) 99
 [Erratum-ibid.\  B {\bf 530} (2002) 260]
 
\bibitem{Guonew}
  F.~K.~Guo, C.~Hanhart and U.~G.~Meissner,
  Eur.\ Phys.\ J.\  A {\bf 40} (2009) 171

  
  
\bibitem{lutzmed}
  M.~F.~M.~Lutz and C.~L.~Korpa,
  Phys.\ Lett.\  B {\bf 633} (2006) 43
  
\bibitem{mizutani}
  T.~Mizutani and A.~Ramos,
  Phys.\ Rev.\  C {\bf 74} (2006) 065201


  
\bibitem{Tolos:2007vh}
  L.~Tolos, A.~Ramos and T.~Mizutani,
  Phys.\ Rev.\  C {\bf 77} (2008) 015207

  
\bibitem{nsd}
  J.~A.~Oller and E.~Oset,
  Phys.\ Rev.\  D {\bf 60} (1999) 074023
  
\bibitem{ollerulf}
  J.~A.~Oller and U.~G.~Meissner,
  Phys.\ Lett.\  B {\bf 500} (2001) 263
  
 
\bibitem{laurap}
  L.~Tolos, A.~Ramos and E.~Oset,
  Phys.\ Rev.\  C {\bf 74} (2006) 015203
  
  

\bibitem{withzou}
  D.~Gamermann, E.~Oset and B.~S.~Zou,
  arXiv:0805.0499 [hep-ph]. Eur.\ Phys.\ J.\  A in print.
  

\bibitem{tolos-dani}
  L.~Tolos, D.~Cabrera, A.~Ramos and A.~Polls,
  Phys.\ Lett.\  B {\bf 632} (2006) 219
  [arXiv:hep-ph/0503009].

  \bibitem{weinberg}
 S.~Weinberg,
 Phys.\ Rev.\  {\bf 130} (1963) 776 
 \bibitem{efimov}
G.~V.~Efimov and M.~A.~Ivanov,
IOP Publishing, Bristol \& Philadelphia (1993)
\bibitem{jnieves}
 H.~Toki, C.~Garcia-Recio and J.~Nieves,
 Phys.\ Rev.\  D {\bf 77}, 034001 (2008)

\bibitem{hanhart}
 V.~Baru, J.~Haidenbauer, C.~Hanhart, Yu.~Kalashnikova and A.~E.~Kudryavtsev,
 Phys.\ Lett.\  B {\bf 586} (2004) 53
\bibitem{Dong}
 Y.~b.~Dong, A.~Faessler, T.~Gutsche and V.~E.~Lyubovitskij,
 Phys.\ Rev.\  D {\bf 77} (2008) 094013 

\bibitem{danibreak}
 D.~Gamermann and E.~Oset,
 arXiv:0905.0402 [hep-ph]. Phys.\ Rev.\  D in print.

\bibitem{Tolos:2004yg}
  L.~Tolos, J.~Schaffner-Bielich and A.~Mishra,
  Phys.\ Rev.\  C {\bf 70} (2004) 025203
\bibitem{tolos}
  L.~Tolos, J.~Schaffner-Bielich and H.~Stoecker,
  Phys.\ Lett.\  B {\bf 635} (2006) 85
  
  

  
\bibitem{ecker}
  G.~Ecker,
  Prog.\ Part.\ Nucl.\ Phys.\  {\bf 35} (1995) 1
  
  
\bibitem{ulfreport}
  V.~Bernard, N.~Kaiser and U.~G.~Meissner,
  Int.\ J.\ Mod.\ Phys.\  E {\bf 4} (1995) 193
  
\bibitem{ulfepj}
  B.~Borasoy, P.~C.~Bruns, U.~G.~Meissner and R.~Nissler,
  Eur.\ Phys.\ J.\  A {\bf 34} (2007) 161
  
  
\bibitem{close}
  F.~E.~Close and R.~G.~Roberts,
  Phys.\ Lett.\  B {\bf 316} (1993) 165
  
\bibitem{borasoy}
  B.~Borasoy,
  Phys.\ Rev.\  D {\bf 59} (1999) 054021
  
\bibitem{PDG}
  W.~M.~Yao {\it et al.}  [Particle Data Group],
  J.\ Phys.\ G {\bf 33} (2006) 1

\bibitem{su4}
  E.~M.~Haacke, J.~W.~Moffat and P.~Savaria,
  J.\ Math.\ Phys.\  {\bf 17} (1976) 2041.
  
\bibitem{angelsfi}
  E.~Oset and A.~Ramos,
  Nucl.\ Phys.\  A {\bf 679} (2001) 616
 
\bibitem{Holinde}
  R.~Machleidt, K.~Holinde and C.~Elster,
  Phys.\ Rept.\  {\bf 149} (1987) 1.

\bibitem{Chanfray}
 G.~Chanfray, D.~Davesne, M.~Ericson and M.~Martini,
 Eur.\ Phys.\ J.\  A {\bf 27} (2006) 191
 
\bibitem{Toki}
 E.~Oset, H.~Toki, M.~Mizobe and T.~T.~Takahashi,
 Prog.\ Theor.\ Phys.\  {\bf 103} (2000) 351


\bibitem{Navarra}
  F.~S.~Navarra, M.~Nielsen and M.~E.~Bracco,
  Phys.\ Rev.\  D {\bf 65} (2002) 037502
\bibitem{waas}
  F.~Klingl, T.~Waas and W.~Weise,
  Phys.\ Lett.\  B {\bf 431} (1998) 254



  
  
\bibitem{cabrera}
  D.~Cabrera and M.~J.~Vicente Vacas,
  Phys.\ Rev.\  C {\bf 67} (2003) 045203

\bibitem{Nieves:1993ev}
  J.~Nieves, E.~Oset and C.~Garcia-Recio,
  Nucl.\ Phys.\  A {\bf 554} (1993) 509

\bibitem{Ramos:1999ku}
  A.~Ramos and E.~Oset,
  Nucl.\ Phys.\  A {\bf 671} (2000) 481


\bibitem{Lakhina:2006fy}
 O.~Lakhina and E.~S.~Swanson,
 Phys.\ Lett.\  B {\bf 650} (2007) 159 
 [arXiv:hep-ph/0608011].


  \bibitem{vanBeveren:2005ha}
 E.~van Beveren, J.~E.~G.~Costa, F.~Kleefeld and G.~Rupp,
 Phys.\ Rev.\  D {\bf 74} (2006) 037501 

\bibitem{vanBeveren:2005pk}
 E.~van Beveren, F.~Kleefeld and G.~Rupp,
 AIP Conf.\ Proc.\  {\bf 814} (2006) 143 
\bibitem{bes}
  M.~Ablikim {\it et al.}  [BES Collaboration],
  Phys.\ Lett.\  B {\bf 607} (2005) 243 
\bibitem{oller}
 J.~A.~Oller and E.~Oset,
 Phys.\ Rev.\  D {\bf 60} (1999) 074023 
\bibitem{hiller}
  W.~Broniowski, W.~Florkowski and B.~Hiller,
  Phys.\ Rev.\  C {\bf 68} (2003) 034911
\bibitem{Rapp1}
 R.~Rapp,
 Nucl.\ Phys.\  A {\bf 782} (2007) 275 

\bibitem{Rapp2}
 R.~Rapp and J.~Wambach,
 Adv.\ Nucl.\ Phys.\  {\bf 25} (2000) 1 

\bibitem{imai}
  T.~Ishikawa {\it et al.},
  Phys.\ Lett.\  B {\bf 608} (2005) 215
  


\bibitem{rocafi}
  D.~Cabrera, L.~Roca, E.~Oset, H.~Toki and M.~J.~Vicente Vacas,
  Nucl.\ Phys.\  A {\bf 733} (2004) 130
  

\bibitem{FAIR}
http://www.gsi.de/fair/index.html




\end{thebibliography}
 \end{document}